# Breaking the material-limited temperature coefficient of resistance via carrier feedback in a single transistor

Jiazhen Chen[1#], Yihao Song[1#], David Alexander Montealegre[2], Mingyang Cai[1], Minjoo Larry Lee[2,3], Fengnian Xia[1]*

[1]*Department of Electrical and Computer Engineering, Yale University, New Haven, CT 06511, USA.*

[2]*The Grainger College of Engineering, Department of Electrical and Computer Engineering and Nick Holonyak, Jr. Micro and Nanotechnology Laboratory, University of Illinois Urbana-Champaign, Urbana, IL 61801, USA*

[3]*The Grainger College of Engineering, Department of Materials Science and Engineering, University of Illinois Urbana-Champaign, Urbana, IL 61801, USA*

*Corresponding author. Email: fengnian.xia@yale.edu

#These authors contributed equally to this work.

**Abstract:**

The temperature coefficient of resistance (TCR) is one of the most fundamental properties of a material. Semiconductor materials exhibiting high TCR are promising candidates for applications in high-resolution thermal imaging for autonomous systems, high-precision temperature sensing, and neuromorphic computing. However, the TCR magnitude is typically below 5%/K near 300 K for thermal imaging materials, such as vanadium oxide and amorphous silicon. Inspired by the distinctive characteristic of feedback in electronic circuits, we demonstrate a voltage-tunable TCR of up to 150%/K near 300 K in a two-terminal InGaAs/InP n-p-n transistor, enabled by an internal coherent carrier feedback mechanism. In this device, current amplification arises from a synergistic interplay between temperature-dependent transistor gain and avalanche multiplication. Carriers amplified at the emitter-base junction via the transistor effect are injected into the collector-base junction, where avalanche multiplication generates additional carriers. These excess carriers are then fed back to the emitter-base junction, triggering further transistor amplification. This regenerative positive feedback loop results in a high and bias-tunable temperature coefficient of resistance (TCR). This work reveals the potential of device engineering in overcoming the fundamental material-level physical limits of temperature properties.

**Introduction**

Thermal sensing and imaging are playing increasingly important roles in intelligent systems[1-6]. Thermal imaging can function in low visibility by collecting emission from objects in the long wavelength infrared (LWIR) band, 8 to 12 µm. The thermal images can then be interpreted together with visible (RGB) images by using various artificial intelligence models to improve the understanding of the environment that autonomous systems such as robots, drones, and self-driving vehicles will navigate[4]. In the LWIR spectral range, cryogenic mercury cadmium telluride (MCT) "photon" detectors can directly convert the low-energy photons into electron-hole pairs that can be detected with high sensitivity and speed, but their high-cost cryogenic operating temperature prevents their wide deployment[7-9]. In contrast, uncooled LWIR detectors have many advantages, including compactness and room-temperature operation. Among the types of uncooled LWIR detectors, microbolometers have been the most widely deployed, in which detection is realized by measuring the variation of material resistance due to LWIR-induced heating[10-12]. However, their performance significantly lags compared to cooled "photon detectors" due to low temperature coefficient of resistance (TCR), long thermal time constant, and high noise. Out of these factors, the TCR is essential since it is directly proportional to the signal generated by the LWIR light. The TCR, defined as $\frac{1}{R}\frac{dR}{dT}$, is $-\frac{E_a}{kT^2}$, where $R$ is the resistance, $E_a$ is the activation energy, $k$ is the Boltzmann constant, and $T$ is the temperature in Kelvin[13-15]. The magnitude of TCR in widely used vanadium oxide and amorphous silicon is usually below 5% per Kelvin. Although an increase in $E_a$ can improve the TCR, it also results in an exponential rise in device resistance ($R$ is proportional to $\exp\left(\frac{E_a}{kT}\right)$), underscoring the need for alternative strategies to enhance the TCR without compromising conductivity.

In this work, we apply a circuit concept to a semiconductor device to overcome the fundamental TCR limit, which is a classic material property. In a two-terminal n-p-n transistor, we show that a positive feedback mechanism analogous to that in an electronic circuit[16] can enhance the sensitivity of the overall gain. As a result, TCR becomes a design parameter, which can be controlled by tuning the feedback factor through bias voltage. Such a paradigm shift from materials optimization to device engineering opens up a new pathway for designing the temperature properties of semiconductor devices.

**Gain feedback mechanism in transistor**

It is known that in transistors under bias, two types of amplification - classic transistor gain and avalanche multiplication – can coexist and lead to a high total gain[17-20]. Figure 1a illustrates this concept using the band diagram of an indium phosphide (InP)/indium gallium arsenide (InGaAs) n-p-n transistor under forward bias with a floating base, commonly referred to as a phototransistor. In this phototransistor, the emitter, base, and collector are made from n-doped InP, p-doped InGaAs, and n-doped InGaAs, respectively. At the emitter/base interface, the accumulation of holes at the valence band induces the injection of electrons from the emitter, leading to transistor current gain, denoted as $h_{FE}$. These electrons will be accelerated by the electric field in the base-collector junction. If the avalanche threshold energy is reached, avalanche multiplication ($M$) can occur, and electron-hole pairs will be generated. The holes generated by avalanche will be swept back to the emitter/base interface, which leads to further injection of electrons from the emitter. Such a process forms positive feedback with a feedback factor $\beta = M - 1$, since only the extra holes generated by the avalanche multiplication are "fed" back. Moreover, the transistor gain, $h_{FE}$, corresponds to the open-circuit gain $A$ in a classic circuit with feedback, as illustrated in Fig. 1b.

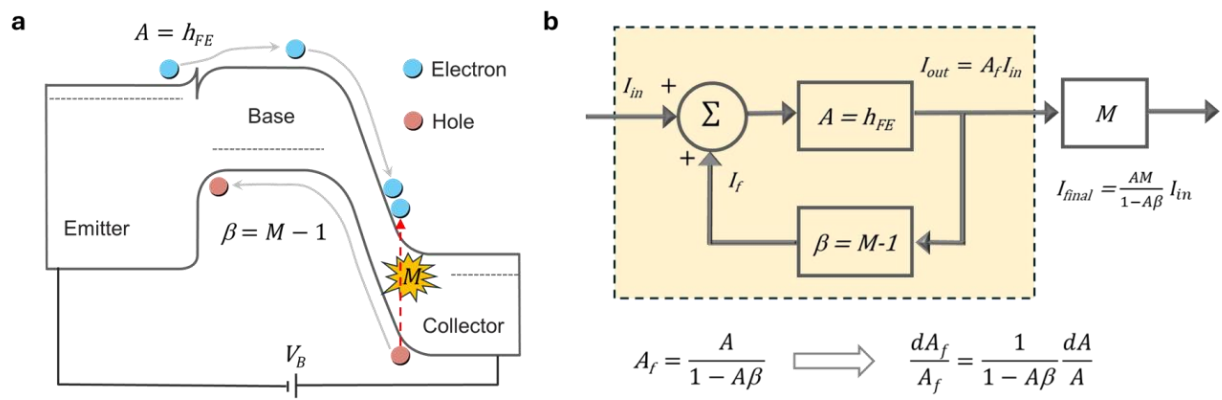

**Figure 1. Gain feedback mechanism in two-terminal n-p-n transistor.** (**a**) The band diagram of a biased InP/InGaAs n-p-n transistor. Two amplification mechanisms are shown. (**b**) Depiction of the contribution of dual amplification mechanisms to the overall gain. The component of hole current in final output is ignored for simplicity.

In Fig. 1b, $I_{in}$ represents the unamplified current, which can be the unamplified diffusion current of the reverse biased base-collector junction due to thermal excitation, the unamplified photocurrent if under light excitation, or the combination of both. If there is no avalanche multiplication ($M = 1$), the device functions as a classic heterostructure transistor with gain of $h_{FE}$ [18,21]. Since in heterostructure transistors $h_{FE} \gg 1$ [22], the total current is dominated by amplified current and the direct contribution of $I_{in}$ in the output can be ignored. When the bias voltage is high enough to induce avalanche, the gain with feedback ($A_f$) is:

$$A_f = \frac{A}{1 - A\beta} = \frac{h_{FE}}{1 - (M-1)h_{FE}}. \tag{1}$$

Eventually, the current resulting from this feedback process will undergo another round of avalanche amplification such that the overall gain, $A_{final}$, is $\frac{AM}{1-A\beta}$. $A_{final}$ can be approximated to $A_f$ since $M$ is kept close to 1 in order to operate the device below the breakdown. In classic circuits, negative feedback is widely utilized to reduce the gain sensitivity by a factor of $\frac{1}{1+A\beta}$ [16]. In the case of positive feedback, the gain sensitivity is enhanced by a factor of $\frac{1}{1-A\beta}$. Since this enhancement of gain sensitivity is universal and independent of the varying sources, such a one-transistor "positive feedback circuit" is also expected to have strong temperature dependence, which can be further tuned by $\beta$.

**Overcoming the TCR limit by feedback in dark conditions**

Figure 2a shows the temperature-dependent I-V characteristics of an InP/InGaAs n-p-n phototransistor in dark conditions. The device is about 200 μm by 400 μm, and the base is floating. The structure of the phototransistor and the biasing scheme are shown in the inset of Fig. 2b. The structure was grown on a heavily n-doped InP substrate at a concentration of $2\times10^{18}$ cm$^{-3}$. First, a 1-μm thick InP layer (emitter) doped with silicon at $1\times10^{18}$ cm$^{-3}$ was grown, followed by 1.5-μm (base) and 0.5-μm (collector) thick In$_{0.53}$Ga$_{0.47}$As layers doped with beryllium and silicon at $2.5\times10^{16}$ cm$^{-3}$ and $1\times10^{17}$ cm$^{-3}$, respectively. Finally, a 50-nm thick highly n$^+$-doped In$_{0.53}$Ga$_{0.47}$As layer was utilized to cap the structure for minimal metal contact resistance. Details of the growth, device fabrication, and characterization are presented in the Supplementary Information.

The device exhibits strong temperature dependence. First, as illustrated in Fig. 2b, the breakdown voltage reduces from 1.79 V to 1.45 V when the temperature increases from 289.9 K to 304.9 K; in the measurement, a temperature detector was placed beside the phototransistor to accurately monitor the operational temperature. Here, the breakdown is due to the synergistic effect of transistor amplification and avalanche multiplication. As illustrated in Eq. (1), breakdown occurs when $M = 1 + \frac{1}{h_{FE}}$, which is only slightly larger than unity due to the high transistor gain. As a result, the breakdown voltage here is much lower than that in traditional InP/InGaAs avalanche photodetectors[23-26]. The decrease in the breakdown voltage is due to the increase of the transistor gain as the temperature rises. Second, the current exhibits strong temperature dependence, especially when the device is biased close to breakdown. Fig. 2c shows the device resistance as a function of temperature for different bias ($V_B$) from 0.6 V to 1.45 V. At $V_B$ of 1.45 V, the transistor resistance decreases from around 480 kΩ at 289.9 K to 17 kΩ at 304.9 K. Furthermore, the TCR is calculated based on the I-V measurements shown in Fig. 2a using $TCR(T_2) = \frac{R(T_2) - R(T_1)}{T_2 - T_1} \times \frac{1}{R(T_2)}$, where $R(T_2)$ and $R(T_1)$ represent the resistance at two adjacent temperatures of $T_2$ and $T_1$ ($T_2 > T_1$), respectively. Since at higher temperatures, the device shows smaller resistance, the TCR is negative. The TCR as a function of bias is shown in Fig. 2d for different temperatures. As illustrated in Fig. 2c, the resistance variation as a function of temperature is rather gradual, and the estimated TCR based on the resistance variation of two adjacent temperatures (with a difference of 2 Kelvin) is quite accurate. In fact, calculations based on a temperature difference of 4 Kelvin only introduce a difference of less than 10%.

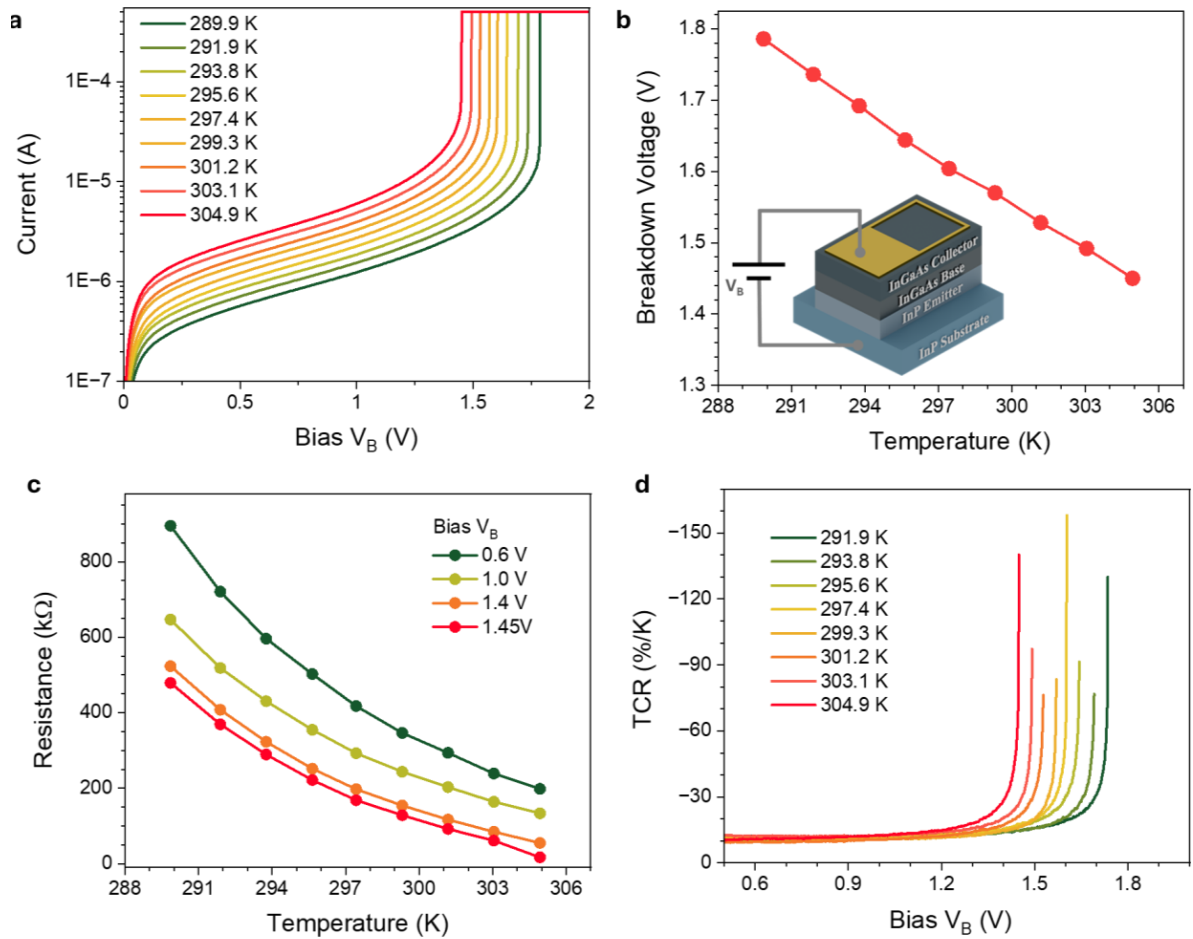

**Figure 2. Two-terminal transistor TCR in dark.** (**a**) The transistor IV characteristics at different temperatures. (**b**) The breakdown voltage vs. temperature. Inset: the schematic of the device and biasing scheme. (**c**) The measured resistance vs. temperature at four biases ranging from 0.6 V to 1.45 V. (**d**) The TCR derived from results in (**a**) at different temperatures.

As illustrated in Fig. 2d, regardless of the operational temperature, the TCR remains largely unchanged (~-11%/K) when the bias is below 0.9 V, and its magnitude increases with bias. Close to the breakdown, a peak TCR of around -150%/K is observed at 297.4 Kelvin. At different temperatures, the measured peak TCR varies from -80%/K to -150%/K due to the temperature-dependent breakdown voltage. Since the TCR beyond breakdown is ill-defined, the accessible TCR depends on how close the device can be biased to breakdown and, in principle, there is no upper limit. In our measurements, we varied the bias voltage at a step of 2 mV with a breakdown compliance current of 0.5 mA. As a result, the measured maximal TCR at different temperatures fluctuates as the breakdown voltage varies with temperature. Away from breakdown, there is no avalanche multiplication, and the total current $I_C$ can be expressed as $I_C \approx h_{FE} I_{CBO}$, where $I_{CBO}$ is the saturation current of the reverse-biased base-collector junction[17,22]. In this case, the TCR depends on the temperature dependence of both $h_{FE}$ and $I_{CBO}$:

$$- TCR = \frac{1}{I_C} \frac{dI_C}{dT} \approx \frac{1}{h_{FE}} \frac{dh_{FE}}{dT} + \frac{1}{I_{CBO}} \frac{dI_{CBO}}{dT} \qquad (2)$$

The temperature dependence of $I_{CBO}$ is largely determined by the bandgap of InGaAs, which is estimated to be around -8%/K. As a result, the temperature dependence of the transistor gain contributes about -3%/K of the measured TCR. As discussed in the next section, our photocurrent measurements confirm the contribution of the temperature dependence of the transistor gain on TCR. Moreover, a similar temperature dependence of gain was reported previously in silicon transistors[27,28]. As bias increases, $M$ will be larger than 1, but it will remain very close to 1 before breakdown (e.g., if $h_{FE} = 500$, $M$ is always between 1 and 1.002 before breakdown). Taking both the avalanche amplification and feedback into account, $I_C \approx \frac{h_{FE}}{1-(M-1)h_{FE}} I_{CBO}$ as illustrated in Eq. (1). In the analysis of the temperature dependence of the total current with feedback, the temperature dependence of $M$ can be ignored since it is always very close to 1 regardless of the operational temperature in this work. In fact, the temperature dependence of breakdown voltage reported in Fig. 2b further indicates that reduction in the breakdown voltage at higher temperature is due to the increase of the transistor gain, and such temperature dependence is distinctively different from that of conventional avalanche photodetectors, in which avalanche breakdown voltage alone increases at higher temperature. The TCR with avalanche feedback is evaluated to be:

$$- TCR = \frac{1}{I_C}\frac{dI_C}{dT} \approx \frac{1}{1-(M-1)h_{FE}}\frac{1}{h_{FE}}\frac{dh_{FE}}{dT} + \frac{1}{I_{CBO}}\frac{dI_{CBO}}{dT} \qquad (3)$$

Compared to the TCR without feedback denoted in Eq. (2), as the device operates with feedback, the first term on the right-hand side of Eq. (3) can be the dominant factor. As a result, the TCR increases when the bias voltage is high enough to induce the avalanche and feedback, as illustrated in Fig. 2d. Most importantly, such a voltage tunable TCR allows the device to operate over a broad temperature range, since the TCR can be tuned by bias and it can be adjusted when the operational temperature varies.

**Temperature dependence of gain**

To clarify the temperature dependence of the transistor gain, we further characterized the photoresponse with 1.55 μm light excitation at different temperatures. In this measurement, light modulated with a mechanical chopper is utilized to excite the phototransistor, and amplified photocurrent is collected by a lock-in amplifier. This scheme allows us to investigate the temperature dependence of the transistor gain by excluding the dark current, since the lock-in only detects the modulated signal induced by light. Measurement details are included in the Supplementary Information. We emphasize that the photoresponse measurements presented here are not intended to demonstrate long-wavelength infrared (LWIR) detection, but rather to elucidate the intrinsic temperature dependence of the transistor gain. To enable LWIR detection, appropriate LWIR-absorbing materials or structures must be integrated with the device in this work.

Fig. 3a shows the photocurrent as a function of bias voltage in the range of 0.4 V to 0.8 V at different temperatures. Within this low bias voltage range, the avalanche does not occur since the threshold energy is not reached [29]. At higher bias, both the depletion width of the base-collector junction and the field within it increase, leading to enhanced collection of photocarriers and larger photoresponse. Moreover, at higher temperatures the photoresponse also increases, indicating higher gain. Fig. 3b summarizes the photoresponses as a function of temperature for four biases ranging from 0.44 V to 0.74 V. At a bias of 0.74 V, the extrinsic photoresponsivity increases from around 520 A/W to 1140 A/W from 281 K to 303 K. Such an enhancement at higher temperature can be due to the following reasons. First, increased temperature facilitates the hole generation in the collector-base junction by heat, and these holes can accumulate in the base-emitter junction to induce electron injection from the emitter[21]. Second, the increased temperature reduces the required forward bias voltage for the emitter-base junction, leading to increased reverse-biased base-collector voltage at a given bias $V_B$. This effect increases the depletion width of the base-collector junction and reduces the effective base width, leading to enhanced current due to electron diffusion in base, which is the dominant source of current in n-p-n transistors. As illustrated in Fig. 3b, at 295 Kelvin, the photocurrent increases by around 3% to 4%/K depending on the bias voltage.

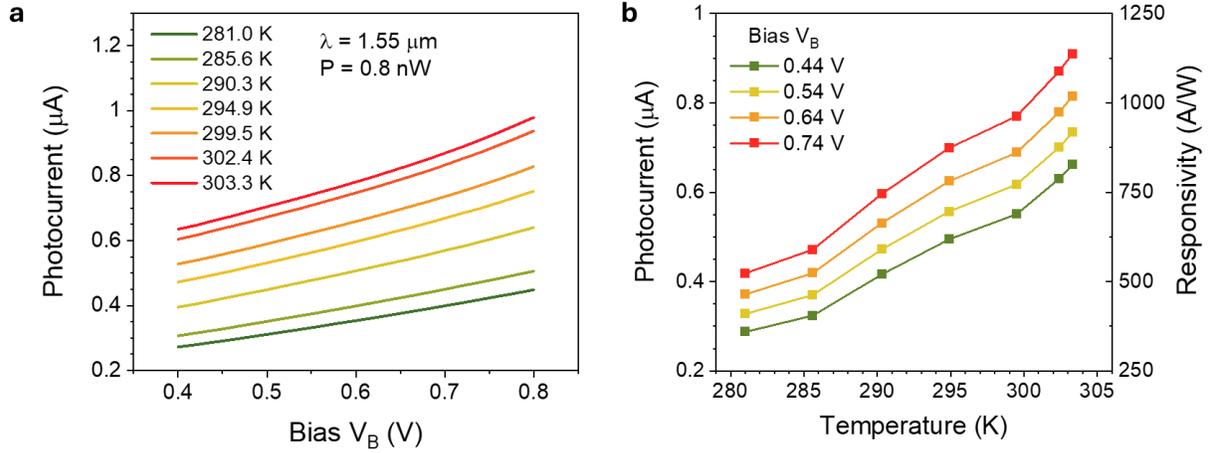

**Figure 3. Photoresponse without feedback effect.** (**a**) The photoresponse as a function of bias voltage measured at different temperatures. The optical power on the device is 0.8 nW. (**b**) Photoresponse as a function of temperature measured at different bias voltages.

We further characterized the photoresponse and its temperature dependence at large bias voltage, in which the avalanche feedback plays a role. Fig. 4a illustrates the photocurrents measured at different temperatures at bias from 1.2 V to breakdown. Beyond the breakdown voltage, the lock-in amplifier could no longer detect the photoresponse signal, and the measurements were discontinued. The measurements were performed separately from those in Fig. 3a, because the photocurrents were much larger, and the measurement parameters were different, as detailed in the Supplementary Information. The calculated temperature coefficient (TC) of photocurrent is calculated using $TC(T_1) = \frac{I(T_2) - I(T_1)}{T_2 - T_1} \times \frac{1}{I(T_1)}$ and plotted in Fig. 4b. When the bias voltage is close to the breakdown, the temperature coefficient is consistently larger than 100%/K. This observation further confirms that the feedback mechanism dominates the high TCR shown in Fig. 2d.

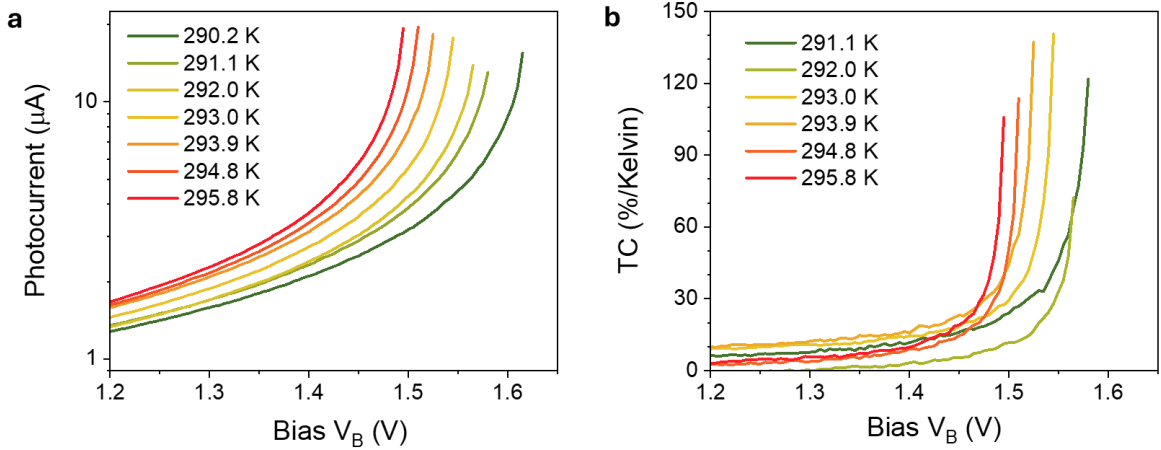

**Figure 4. Photoresponse with feedback.** (**a**) The photoresponse as a function of bias voltage measured at different temperatures. The optical power on device is 0.8 nW. (**b**) The extracted temperature coefficient (TC) as a function of the bias voltage at different temperatures.

## Conclusions

In summary, we have applied the feedback concept in electronic circuits to a two-terminal floating-base transistor to overcome the material-imposed limit on TCR. While this principle is demonstrated using a heterostructure InP/InGaAs transistor, the approach can be broadly applicable to transistors made from any semiconductor materials. This concept thus creates substantial opportunities to design transistor-based bolometers with tailored TCR, resistance, and operating current or voltage to suit specific applications. Future integration of LWIR absorption materials or structures with feedbacked semiconductor phototransistors will enable more sensitive and tunable thermal detection. The transistor-based bolometers can improve the accuracy and efficiency of thermal vision, environment mapping, and intelligent monitoring, expanding the capabilities of AI in complex real-world conditions.

## Acknowledgements

We thank Mark Han, Romil Audhkhasi, Haoqing Deng, Virat Tata, Arka Majumdar, and Mo Li at University of Washington, Seattle and Wenjun Deng and Qiushi Guo at Research Foundation of CUNY- Advanced Science Research Center for insightful discussions. We acknowledge the use of ChatGPT Plus (OpenAI) to assist with language refinement and sentence structuring in the preparation of this manuscript. **Funding**: this work is supported by the Office of Naval Research (N000142412307) and Yale University. Epitaxial wafers grown with support from the Defense Advanced Research Projects Agency (HR00112290005) were utilized in this work.

## Competing Financial Interests

A non-provisional patent application on optical sensing based on the gain feedback was submitted to USPTO in March 2025.